          [1999/12/02 1.01 (PWD)]
\documentclass[a4paper,twocolumn]{esapub} 

\def\na{\mbox{$^{22}$\hspace{-0.1em}Na}}            
\def\MeV{\mbox{Me\hspace{-0.1em}V}}                 
\def\Msol{\hbox{M$_{\odot}$}}                       
\def\deg{\hbox{$^\circ$}}                           
\def\funit{\mbox{photons cm$^{-2}$ s$^{-1}$}}       
 
\def\py{\mbox{yr$^{-1}$}}
\def\gray{\mbox{$\gamma$-ray}}                      

\usepackage{graphicx}

\title{Upper limits of the $^{22}$Na yield from O-Ne nova}
\author[1]{P. Jean}
\author[1]{J. Kn\"odlseder}
\author[1]{P. von Ballmoos}
\affil[1]{CESR, UPS-CNRS, Toulouse, France}
\author[2]{J. G\'omez-Gomar}
\author[2]{M. Hernanz}
\author[2,3]{J. Jos\'e}
\affil[2]{IEEC, Edifici Nexus, C/Gran Capit\`at 2-4, 08034 Barcelona, Spain}
\affil[3]{Dpt. F\'{\i}sica i Enginyeria Nuclear, Av. V\'{\i}ctor Balaguer, E-08800 Vilanova i la Geltr\'u (Barcelona), Spain}

\begin{document}


\maketitle

\begin{abstract}
The radioactive isotope $^{22}$Na is one of the primary tracer isotopes that may potentially be detectable by gamma-ray spectroscopy. This isotope is predicted to be produced in ONe nova explosions, yet no detection of its 1.275 MeV gamma-ray signature has been reported so far.
In particular, 6 years of COMPTEL observations of the galactic bulge, a region that should be particularly rich in nova explosions, have only led to an upper-limit. In this paper we will present the implications of the COMPTEL upper flux limit on the permissible $^{22}$Na yield for ONe novae. We base our analysis on extensive Monte-Carlo simulations of galactic 1.275 MeV emission that were constrained by the most recent results of galactic nova rates and distributions. We demonstrate that the non-detection of the 1.275 MeV line from the galactic bulge implies a solid upper $^{22}$Na yield limit in agreement with current theoretical nucleosynthesis calculations.
\end{abstract}

\section{Introduction}
Classical ONe nova outbursts are expected to synthesize \na\ and provide a \gray\ signal at 1.275~\MeV\ (Clayton \&\ Hoyle (1974), G\'omez-Gomar et~al. (1998)). Observations with several instruments have reported only upper-limits on the 1.275~\MeV\ flux from the Galaxy (HEAO3, SMM, OSSE, COMPTEL) or from individual novae (COMPTEL) and consequently allow to derive an upper-limit of the \na\ mass ejected per ONe novae. Observations with the HEAO3 \gray\ spectrometer provided an upper-limit of 4~10$^{-4}$~\funit\ on the Galactic accumulated flux leading to an upper-limit of the ejected \na\ mass of 3~10$^{-7}$~\Msol\ (Higdon \& Fowler, 1987). Leising et al. (1998) found a 99\%\ confidence limit of 1.2 10$^{-4}$~\funit\ on a steady 1.275~\MeV\ flux from the Galactic center (GC) direction. Iyudin et al. (1995), using COMPTEL observations of single novae, estimated a 2$\sigma$ upper-limit of 3 10$^{-5}$~\funit\ for any neon-type novae in the galactic disk, which has been translated into an upper-limit of the ejected \na\ mass of 3.7~10$^{-8}$~\Msol. Using observation of the GC with OSSE, Harris (1997) estimated an upper-limit of the \na\ mass of 2~10$^{-7}$~\Msol. Recently, Iyudin et al. (1999) derived a 2$\sigma$ upper-limit of the ejected \na\ mass of 3.6~10$^{-9}$~\Msol\ and 2.1~10$^{-8}$~\Msol\ from the COMPTEL observations of the bulge and Nova Cygni 1992, respectively. 
However, most of the \na\ mass upper-limits derived from the observation of the Galactic diffuse emission have been calculated using overestimated ONe nova frequencies (up to 40 ONe novae per year only in the bulge). Current estimations of the ONe nova frequency range from 3~\py\ to 15~\py\ in the whole Galaxy (see details of these figures in section 2). Moreover, the flux at 1.275~\MeV\ in the GC region depends not only on the amount of \na\ ejected per outburst and the rate of Galactic ONe novae but also on their spatial distribution. 

In the presented work, we estimate the upper-limit of the ejected \na\ mass derived from COMPTEL observation of the cumulative 1.275~\MeV\ emission in the GC region taking into account recent results of rates and spatial distributions of novae in our Galaxy. We modeled the Galactic emission at 1.275~\MeV\ as a function of the ONe novae rate and the mean \na\ yield per outburst. This has been done for currently used spatial distributions of novae in the Galaxy. The upper-limit of the ejected \na\ mass is estimated as a function of the Galactic ONe nova frequency. The next section describes the method used to model the Galactic emission at 1.275~\MeV. The estimation of the \na\ mass upper-limit derived from the COMPTEL measurement is presented in section 3. Discussion and conclusions follow (section 4).

\section{Modeling of galactic 1.275~\MeV\ emission}
We have used a method based on the work from Higdon \& Fowler (1987) which estimated the first \na\ mass upper-limit using observation of diffuse 1.275~\MeV\ emission with HEAO3. The distribution of the 1.275~\MeV\ emission from galactic ONe novae is simulated with a Monte-Carlo method. The position in galacto-centric coordinates and the age of ONe novae are chosen randomly according to the appropriate distributions and rates. For a given value of the \na\ yield per outburst, the 1.275~\MeV\ flux arriving on Earth from each nova is calculated with the distance and the age of the nova (see Jean et al. (2000) for a detailed description of the method). 

For the purpose of this work, a Galaxy model with a disk and a spheroid is convenient to simulate the distribution of Galactic ONe nova events. Several laws for the spatial distribution of classical novae in the disk and in the `spheroid' (representative of the bulge) have been proposed. We have selected four models that differ significantly from each other. The first and older of them is described in Higdon \& Fowler, 1987 - hereafter HF87. The disk and spheroid models are derived from the starlight surface brightness distribution. By scaling the nova rate in the bulge of M31 to the bulge of our Galaxy, HF87 estimate the proportion of novae in the spheroid to 0.348. 
The second model has been used by Hatano et al. (1997) to estimate the 
occurrence rate of Galactic classical novae and the fraction of novae 
in the bulge. It is based on a model of the distribution of type Ia supernovae by Dawson \& Johnson (1994) - hereafter DJ94. The proportion of novae that 
occur in the bulge is set to 0.111 on the basis of an estimate of the bulge to total galaxy mass ratio. The third model is derived by Kent, Dame \& Fazio (1991) - hereafter KDF91 - from the Galactic survey of the Spacelab InfraRed Telescope (IRT) that provides a reliable tracer of the distribution of G and K giant stars. Using the total infrared luminosity of the bulge and the disk, the derived proportion of novae occurring in the bulge is 0.179. The last model is taken from Van der Kruit (1990) - hereafter VdK90. It has been used by 
Shafter(1997) to estimate the galactic nova rate. This author assumes 
that the nova distribution follows the brightness profile of our 
Galaxy. Under this assumption, the proportion of bulge novae is 0.105. Table \ref{tab:model} summarizes the characteristics of the adopted models.

\begin{table} 

\caption{Characteristics of the models adopted for the simulations of the 
emission at 1.275~\MeV\ from galactic ONe novae. R$_{1/2}$ is the 
half radius of the spheroid, $\rho_h$ the radial scalelength of the disk, 
$z_h$ the scaleheight of the disk and p$_b$ the proportion of novae that 
occur in the bulge.}

\smallskip
\begin{array}[b]{|l|cccc|}
\hline 
{\rm Models} & R_{1/2} (kpc) & \rho_h (kpc) & z_h (kpc) & p_b \\ \hline
{\rm HF87}   & 2.7 &  3.5     & 0.106 & 0.348 \\
{\rm VdK90}  & 2.7 &  5.0     & 0.300 & 0.105 \\
{\rm KDF91} & 1.4 &  3.0     & 0.170 & 0.179 \\
{\rm DJ94}  & 1.4 &  5.0     & 0.350 & 0.111 \\ \hline

\end{array} 


\label{tab:model}
 
\end{table}

The Galactic nova rate is poorly known,
independently of the underlying white dwarf composition, because the
interstellar extinction prevents us from directly observing more than a 
small fraction of the novae that explode each year. Several methods have been 
used to estimate the nova occurrence rate. Estimations based on 
extrapolations of Galactic nova observations suggest a rate in the range 
50-100 \py (see Shafter (1997) and references therein). 
Observations of novae in other galaxies have revealed a 
correlation between the nova rate and the infrared luminosity of the 
parent galaxy. However, the occurrence rates of Galactic novae computed by 
scaling the nova rate measured in external galaxies ($\approx$ 20~\py, see Dawson \&\ Johnson (1994)) are lower than 50 \py. Recently, Shafter (1997) reconciled this 
difference by recomputing the nova rate with the Galactic nova data. He 
extrapolated the global nova rate from the observed one, accounting 
for surface brightnesses of the bulge and the disk components and 
correction factors taking care of any observational incompleteness. With 
this method, Shafter (1997) estimated the nova rate to be 35$\pm$11~\py. Hatano et al. (1997) found a similar value (41$\pm$20~\py) using a Monte-Carlo technique with a simple model for the distribution of dust and classical novae in the Galaxy.

The Galactic ONe nova rate is obtained by multiplying the total nova rate 
by the fraction of novae that results from thermonuclear runaways in 
accreted hydrogen-rich envelopes on an ONe white dwarf. Several authors deduced from observations of abundances in nova ejecta a proportion of ONe novae from 20 to 57 percent of observed nova outbursts. Livio \& Truran (1994) reestimated the frequency of occurrence of ONe novae, in light of observations of abundances in nova ejecta. They concluded that, of the 18 classical novae for which detailed abundance analyses were available, only two or three had a large 
amount of neon and were ONe novae, whereas three other novae showed a modest enrichment in neon, casting doubt on the type of the underlying white dwarf. Under these considerations, they estimated a fraction of Galactic ONe novae between 11 and 33 percent.

\begin{figure}
\centering
\includegraphics[width=1.0\linewidth]{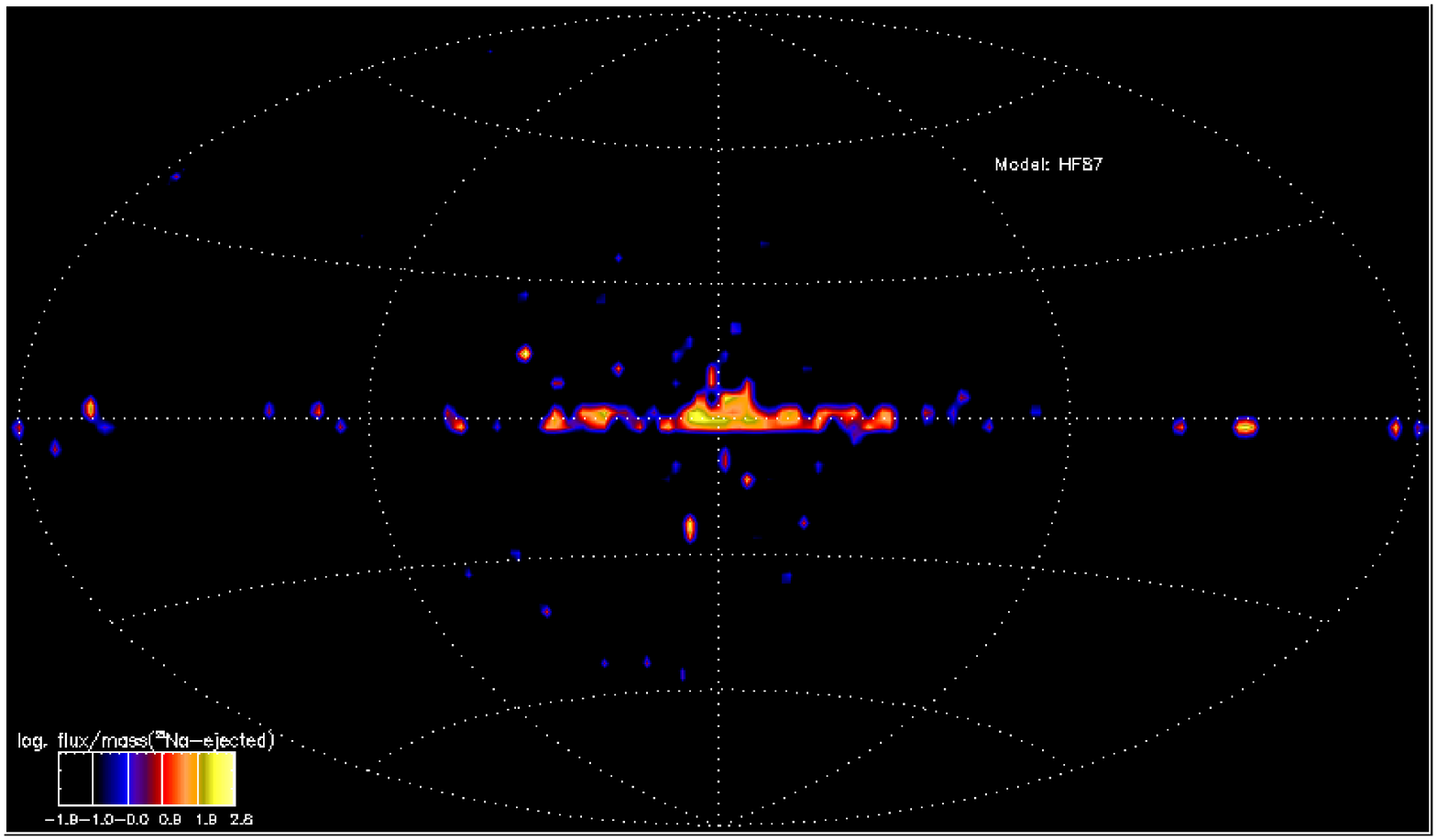}
\includegraphics[width=1.0\linewidth]{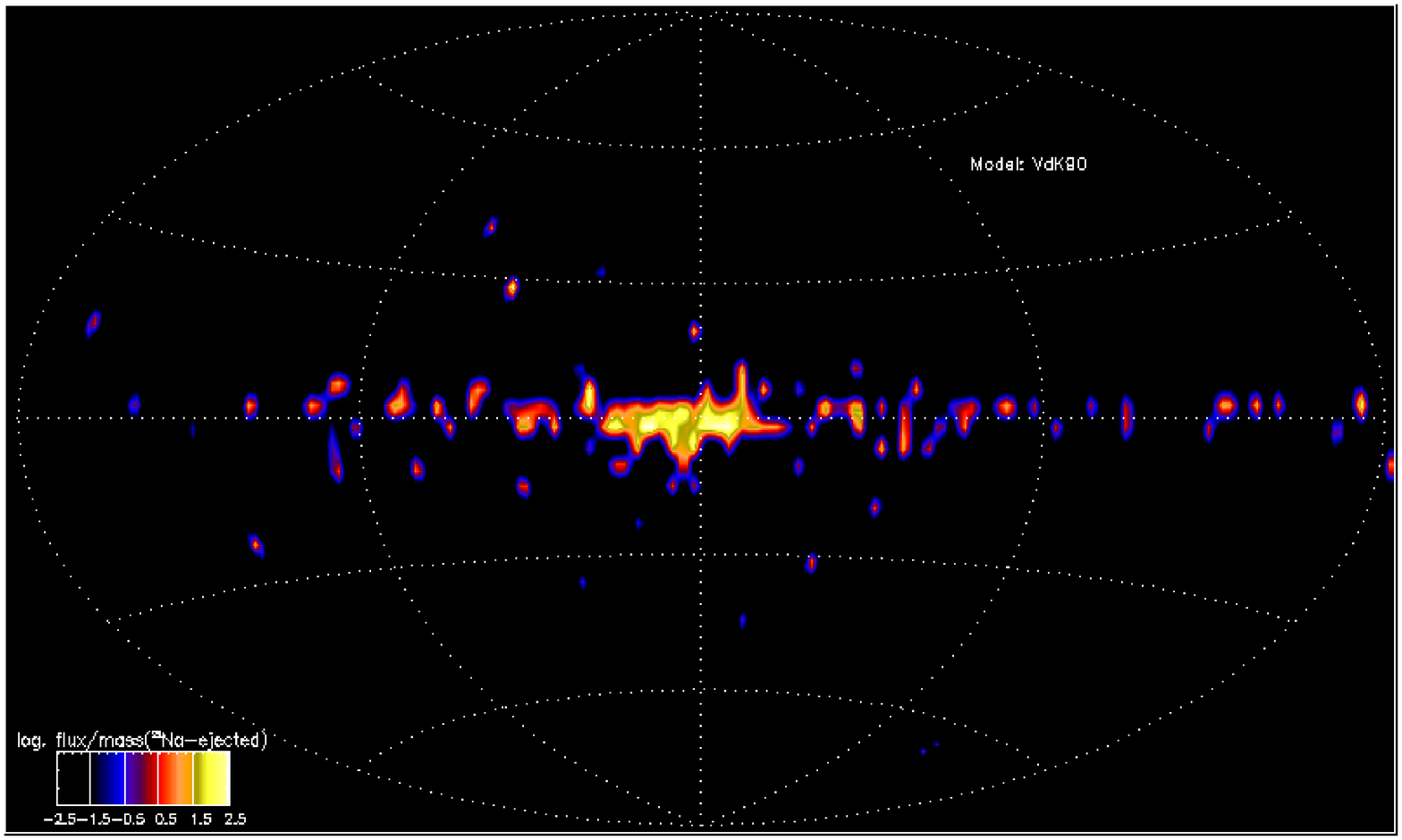}
\includegraphics[width=1.0\linewidth]{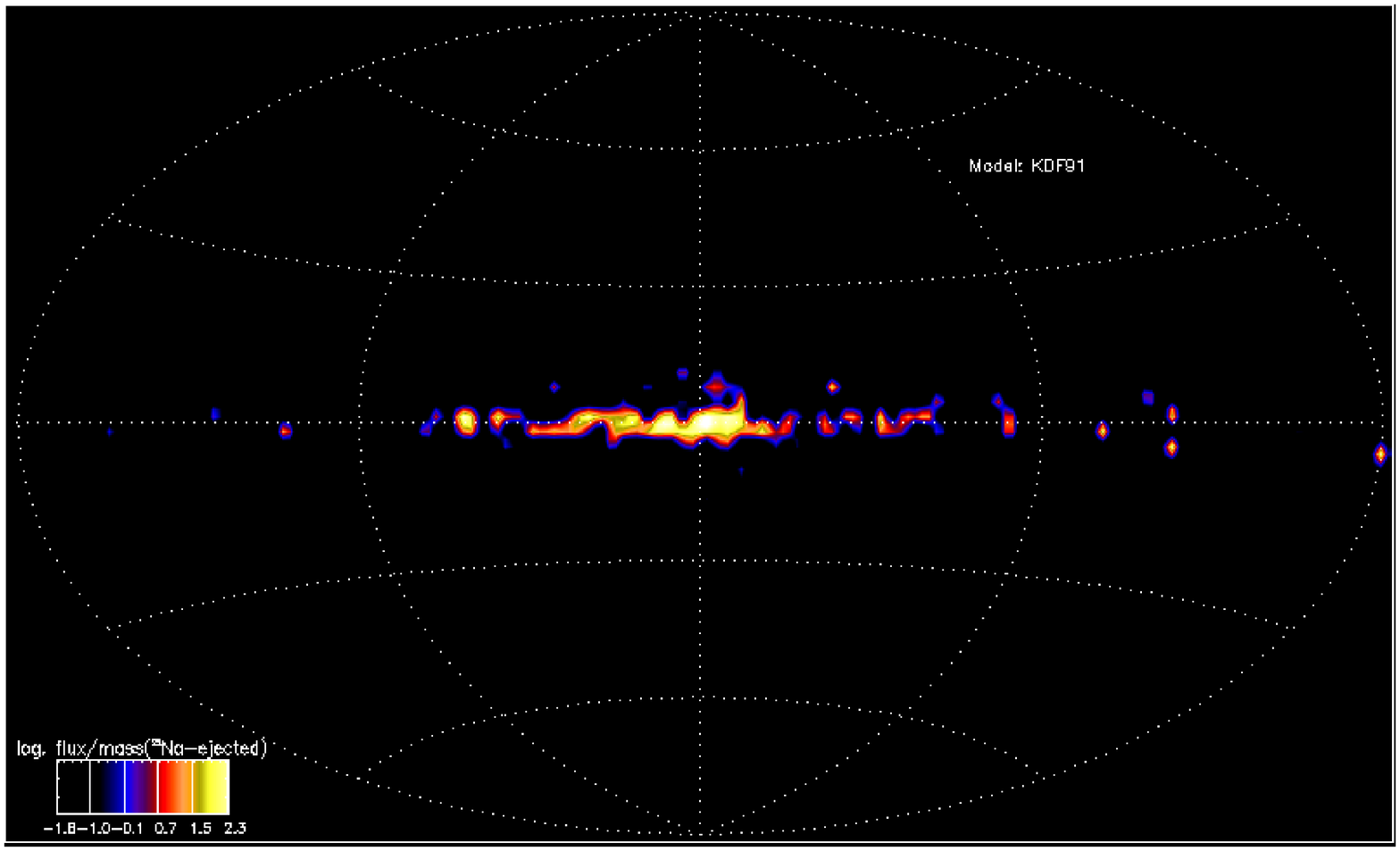}
\includegraphics[width=1.0\linewidth]{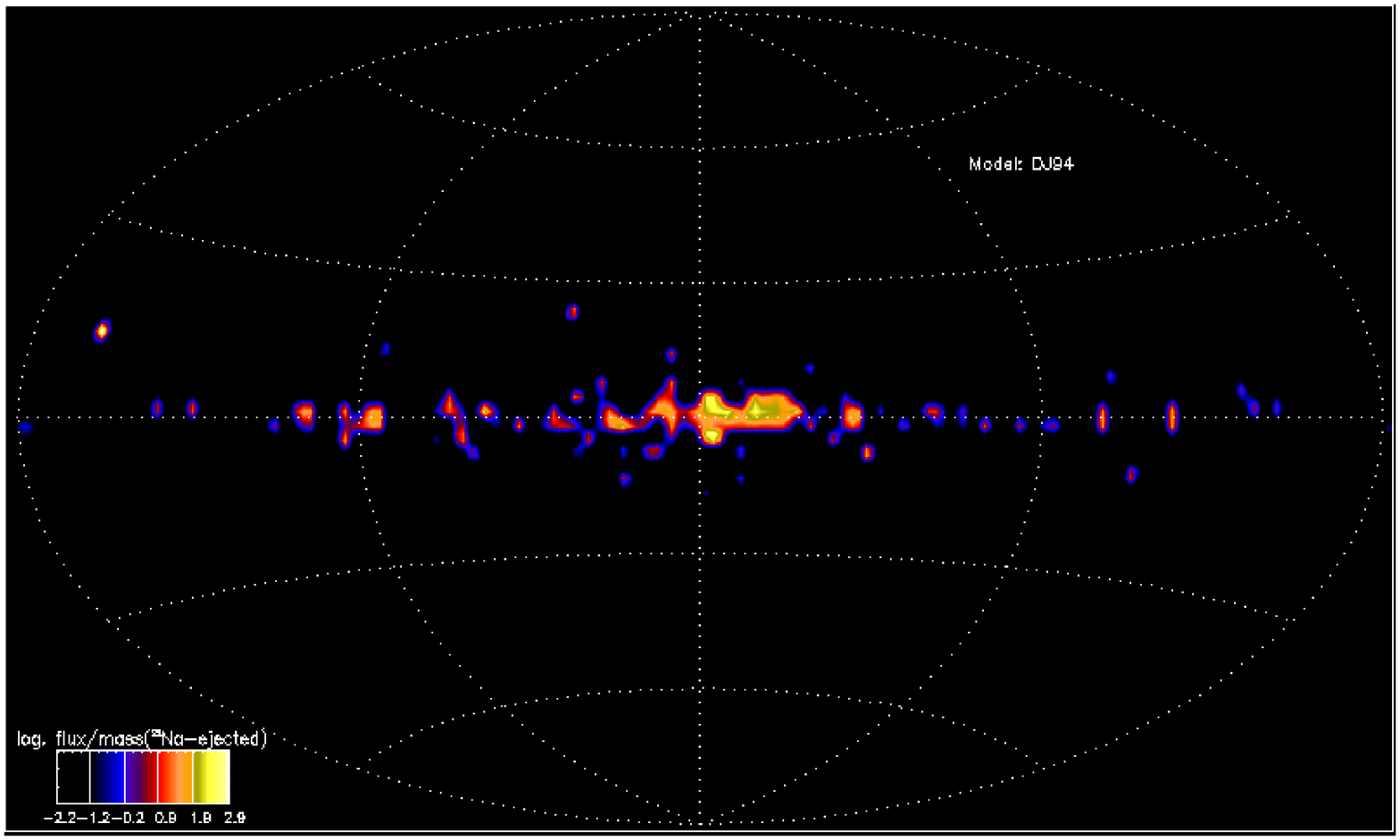}
\caption{Examples of modeled diffuse 1.275 MeV emission generated by Monte Carlo simulation for the four chosen spatial distributions. The adopted ONe nova rate is 10 \py. The fluxes have been normalized with the \na\ ejected mass. The intensity is plotted with 3$^o$ by 3$^o$ pixels in a map in galactic coordinates. \label{figdist}}
\end{figure}

Therefore, for the calculations presented in the next section, we have adopted
frequencies of Galactic novae ranging from 24~\py\ to 46~\py\ and a proportion 
of ONe novae from 11~\%\ to 33~\%, in agreement with the recent
estimates of Livio \& Truran (1994) and Shafter (1997). These values correspond to a lower and upper-limit of the ONe nova rate of 3~\py\ and 15~\py, respectively. Figures \ref{figdist} show examples of computed 1.275 MeV line flux spatial distribution for the four chosen models and a galactic ONe nova rate of 10 \py.

\section{Estimation of the \na\ upper-limit}

\begin{figure}
\centering
\includegraphics[width=1.0\linewidth]{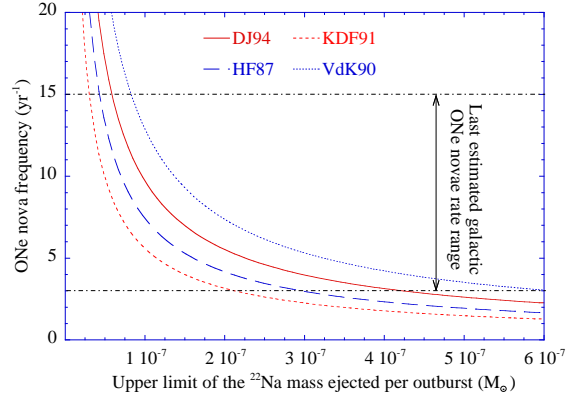}
\caption{Relation between the \na\ mass upper-limit derived from the COMPTEL 1.275~\MeV\ flux upper-limit in the 10\deg\ radius around the GC, and the Galactic ONe nova rate, for the 4 adopted Galaxy models. \label{figure1}}
\end{figure}

Several Monte-Carlo simulations have been done for a given galactic ONe-novae frequency-spatial distribution and a fixed value of the \na\ yield per nova. For a large number N$_T$ of Galaxy-tests, the probability for a detection of the cumulative emission has been estimated by calculating the number N$_i$ of Galaxy-test for which the simulated flux in a circle of 10\deg\ radius around the GC is above the upper-limit flux of COMPTEL ($f_{2\sigma}$=3.0 10$^{-5}$~\funit\ - Iyudin et al., (1999)) during the lifetime (6 years) of the instrument. The probability is obtained by dividing N$_i$ with the total number of Galaxy-test N$_T$. Such fractions above $f_{2\sigma}$ can be interpreted simply as the fraction of time COMPTEL would detect such flux. The \na\ mass upper-limit is defined when the probability of detection of the 1.275~\MeV\ flux by COMPTEL reaches 90\%. It has been calculated for several galactic ONe nova rate and for the proposed spatial distribution models. The results are presented in the figure \ref{figure1}.

\section{Discussion}
\begin{figure}
\centering
\includegraphics[width=1.0\linewidth]{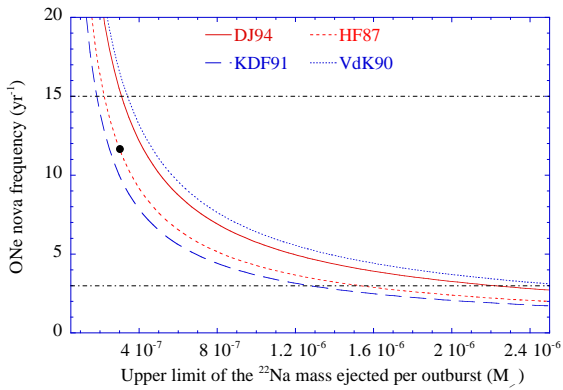}
\caption{Relation between the \na\ mass upper-limit, derived from the HEAO3 Galactic diffuse 1.275~\MeV\ flux upper-limit (in a longitudinal band ranging from 330\deg\ to 30\deg), and the Galactic ONe nova rate, for the 4 adopted Galaxy models. The dot shows the estimation of Higdon \& Fowler (1987). \label{figure2}}
\end{figure}

The \na\ mass upper-limit derived from the 1.275~\MeV\ emission of the bulge depends strongly on the adopted ONe nova rate and spatial distribution. Based on recent results on ONe nova rate, we show that this upper-limit can not be lower than 2~10$^{-8}$~\Msol\ per nova (see figure \ref{figure1}). Moreover, taking into account the uncertainty in the ONe nova frequency in our Galaxy, the most reliable upper-limit should be estimated using the lowest ONe nova rate value and the less favorable spatial-distribution. Consequently, with an ONe nova rate of 3~\py\ and the VdK90 model we obtain a \na\ mass upper-limit of 6~10$^{-7}$~\Msol\ with the COMPTEL measurement. Indeed, if the yield of \na\ is larger than this upper-limit, COMPTEL should have detected the 1.275~\MeV\ line in the bulge especially if the ONe nova rate is larger and if the distribution is more favorable for the detection of this \gray\ line. Applying the same analysis with the 1.275~\MeV\ flux upper-limit derived by the observation of the central 60\deg\ of the Galaxy by the HEAO3 spectrometer (Higdon \& Fowler (1987)) yields a \na\ mass upper-limit of 2.5~10$^{-6}$~\Msol\ instead of 3~10$^{-7}$~\Msol\ (see figure~\ref{figure2}). 

The \na\ mass upper-limit derived by Iyudin et al. (1999) is $\approx$170
times lower than the value obtained from our analysis of COMPTEL measurements. 
The difference can be explained by the different O-Ne nova rate that has been 
assumed for the galactic bulge region.
While Iyudin et al. (1999) assumed 40 O-Ne novae per year in the bulge
region, we estimate this rate to only about 0.3 per year (see table 1 and section 2).
Iyudin et al. (1999) simply employed the 'canonical frequency of 40 novae
\py' in the Galaxy as the O-Ne nova frequency of the galactic bulge region.
However, only about 1/4 of all galactic novae are of O-Ne type, and also,
only a small fraction of them is located within the galactic bulge.
Hence we believe that our more conservative estimate is more realistic,
and the actual constraints imposed by bulge gamma-ray observations on nova \na\
yields are less stringent than claimed by Iyudin et al. (1999).

Since the pioneering work of Clayton \& Hoyle (1974), pointing out the potential role of novae as gamma-ray emitters, several theoretical nucleosynthesis studies have been performed to predict the amount of \na\ ejected in nova outbursts. They have shown that only ONe novae are likely sites for the synthesis of \na\ due to the higher peak temperatures attained during the explosion (with respect to CO novae) but also due to the presence of seed nuclei mainly $^{20}$Ne, extremely relevant to \na\ synthesis since CNO breakout scarcely takes place at typical nova temperatures. Recent fully hydrodynamical calculations of nova outbursts with updated nuclear reaction rates indicate that nova release about 7~10$^{-9}$ (for 1.15~\Msol\ white dwarf) to 4.4~10$^{-9}$ \Msol\ (1.35~\Msol\ white dwarf) of \na\ into the interstellar medium during the explosion (Jos\'e \& Hernanz (1998), Jose, Coc \& Hernanz (1999)). Such values are below the limits posed by several attempts to find the 1.275~\MeV\ gamma-ray signature from classical novae. Although nuclear uncertainties associated to key reactions relevant to \na\ synthesis exist (i.e., $^{21}$Na(p,$\gamma$) and \na(p,$\gamma$)) their incidence in the expected yields is small, namely a factor of 3. A recent reanalysis of the uncertainties associated to $^{21}$Na(p,$\gamma$) (Smirnova \& Coc (2000), Coc et~al. (2000)) suggest even a smaller impact on the \na\ yields.

The most stringent \na\ mass upper-limit obtained so far is presented by Iyudin et al. (1999) who derived a 2$\sigma$ upper-limit of 2.1~10$^{-8}$~\Msol\ from COMPTEL observations of Nova Cygni 1992. This upper-limit still remains above the most optimistic theoretical estimates. Observation with future \gray\ spectrometers should provide more stringent upper-limit.  
Jean et al. (2000) show that the likelihood to detect the Galactic diffuse 1.275~\MeV\ emission with SPI (Spectrometer of the INTEGRAL mission) with only $\approx$10 days of observation is small. On the other hand, they estimate that 80~days of observation of the GC region could already provide constraints to the mean \na\ yield in ONe novae as a function of their frequency-spatial distribution in the Galaxy. Using the most recent calculations of \na\ yields in ONe novae and SPI instrumental background predictions, Hernanz et al. (2000) estimate that the 1.275~\MeV\ line from an individual ONe nova could be detected by the spectrometer of INTEGRAL if its distance is less than 1 to 2~kpc, depending on the uncertainties of nuclear reaction rates. 

Gamma-ray observation of novae would provide information not only on their eruption mechanisms and the nucleosynthesis processes involved in their explosion but also on their distribution and their rate in the Galaxy (e.g. proportion in the bulge, scaleheight in the Galactic disk) since the problem of the interstellar extinction does not appear at this energy range.

\section*{Acknowledgments}
Research partially supported by the research projects ESP98-1348, PB98-1183-C02, PB98-1183-C03. We wish to thank the refere, R. Diehl, for his helpful comments. 


\begin{thebibliography}{}

\bibitem[Clayton \&\ Hoyle (1974)]{CH94}
Clayton, D.D \& Hoyle, F., 1974, ApJ, 187, L101

\bibitem[Coc et~al. (2000)]{Coc00}
Coc, A., Smirnova, N., Jose, J., Hernanz, M. \&\ Thibaud, J.P., 2000, Nuclei in the Cosmos, Nucl. Phys., in press 

\bibitem[Dawson \&\ Johnson (1994)]{DJ94}
Dawson, P.~C. \&\ Johnson, R.~G., 1994, JRASC, 88, 369

\bibitem[Della Valle \&\ Livio (1994)]{DVL94}
Della Valle, M. \&\ Livio, M., 1998, AAP, 286, 786

\bibitem[G\'omez-Gomar et~al. (1998)]{Gom98}
G\'omez-Gomar, J., Hernanz, M., Jos\'e, J. \&\ Isern, J., 1998, MNRAS, 296, 913

\bibitem[Harris (1997)]{Har97}
Harris, J.~M., 1997, Proc. of the Fourth Compton Symposium, AIP conference Proceedings, 410, 1094

\bibitem[Hatano et~al. (1997)]{Hat97}
Hatano, K., Branch, D., Fisher, A. \&\ Starrfield, S., 1997, MNRAS, 290, 113

\bibitem[Hernanz et~al. (1997)]{Her97}
Hernanz, M., G\'omez-Gomar, J., Jos\'e, J., \&\ Isern, J., 1997, Proc. of the 2$^{nd}$ INTEGRAL Workshop, ESA publication division, 382, 47

\bibitem[Hernanz et~al. (2000)]{Her00}
Hernanz, M., G\'omez-Gomar, J., Jos\'e, J., \&\ , Coc A., 2000, this Proceeding

\bibitem[Higdon \& Fowler (1987)]{HF87}
Higdon, J.~C., \&\ Fowler, W.~A., 1987, ApJ, 317, 710

\bibitem[Iyudin et~al. (1995)]{Iyu95}
Iyudin, A.~F., et~al., 1995, A\&A, 300, 422

\bibitem[Iyudin et~al. (1999)]{Iyu99}
Iyudin, A.~F., et~al., 1999, Proc. of the 3$^{nd}$ INTEGRAL Workshop, Astro. Lett. \& Communications, 38, 371

\bibitem[Jean et~al. (2000)]{Jea00}
Jean, P., Hernanz, M., G\'omez-Gomar, J. \&\ Jos\'e, J., 2000, MNRAS, vol. 319, 350

\bibitem[Jos\'e \& Hernanz (1998)]{JH98}
Jose, J. \&\ Hernanz, M., 1998, ApJ, 494, 680

\bibitem[Jos\'e, Coc \& Hernanz (1999)]{JCH99}
Jose J., Coc A. \&\ Hernanz M., 1999, ApJ, 520, 347 

\bibitem[Kent, Dame \&\ Fazio (1991)]{KDF91}
Kent, S.~M., Dame, T.~M. \&\ Fazio, G., 1991, ApJ, 378, 131

\bibitem[Leising et~al. (1988)]{Lei88}
Leising, M.~D., Share, G.~H., Chupp, E.~L. \&\ Kanbach, G., 1988, ApJ, 328, 755

\bibitem[Livio \&\ Truran (1994)]{LT94}
Livio, M. \&\ Truran, J.~W., 1994, ApJ, 425, 797

\bibitem[Shafter (1997)]{Sha97}
Shafter, A.~W., 1997, ApJ, 487, 226

\bibitem[Smirnova \&\ Coc (2000)]{SC00}
Smirnova, N. \&\ Coc, A., 2000, Phys. Rev. C, in press (astro-ph/0008241)

\bibitem[Van der Kruit (1990)]{VdK90}
Van der Kruit, P. 1990, in The Milky Way as a Galaxy, ed. R. Buser \&\ I.R. King (Mill Valley: University Science Books), p. 331

\end{thebibliography}

\end{document}